\documentclass[11pt,a4paper]{article}

\usepackage[utf8]{inputenc}
\usepackage[T1]{fontenc}
\usepackage{amsmath,amssymb}
\usepackage{graphicx}
\usepackage{booktabs}
\usepackage[numbers,sort&compress]{natbib}
\usepackage[hidelinks]{hyperref}
\usepackage[margin=1in]{geometry}
\usepackage{newtxtext}
\usepackage{newtxmath}
\usepackage{array}
\usepackage{multirow}

\newcommand{\fited}{FitED}
\newcommand{\vect}[1]{\boldsymbol{#1}}
\newcommand{\mat}[1]{\mathbf{#1}}

\begin{document}

\begin{center}
{\LARGE \bfseries FitED: A User-Centric, Extensible Software Environment for Robust Peak-Profile and General Functional Data Fitting\par}

\vspace{1em}

{\large Mustafa Mahmoud Aboulsaad\par}

\vspace{0.5em}

{\small Department of Materials Science and Engineering, Solid State Physics, Uppsala University, Box 35, 75103 Uppsala, Sweden\par}

\vspace{0.3em}

{\small \text{mustafa.aboulsaad@angstrom.uu.se; mustafa.aboulsaad@outlook.com}\par}
\end{center}

\vspace{1.5em}

\begin{abstract}
Reliable parameter extraction from experimental data is essential for quantitative analysis across spectroscopy, diffraction, photoluminescence, chromatography, microscopy, and time-resolved measurements. However, nonlinear fitting often remains difficult to reproduce, especially when complex models, correlated parameters, uncertain derived quantities, and user-dependent fitting choices are involved. We present \fited, a Python-based desktop application for nonlinear fitting of one-dimensional scientific data that combines an accessible graphical interface with a transparent and flexible numerical backend. \fited supports conventional peak profiles, including Gaussian, Lorentzian, Pseudo-Voigt, and exact area-normalized Voigt functions, as well as arbitrary user-defined analytical models for broader experimental applications. The software integrates local and global-search-assisted optimization strategies, automated model initialization, repeated stability testing, parameter-correlation analysis, and covariance-based propagation of uncertainty for derived quantities. By combining interactive usability with uncertainty-aware analysis and structured export of fitting results, \fited provides a practical platform for reproducible and interpretable fitting of experimental data. The software is intended to support both routine analysis and advanced model evaluation while preserving the parameter-level control required by experimental researchers.
\end{abstract}


\section{Introduction}

Quantitative model fitting is one of the central computational steps in experimental science. In spectroscopy, diffraction, photoluminescence, Raman scattering, chromatography, and time-resolved measurements, the quantities of scientific interest are often not the raw data points alone, but fitted parameters such as peak positions, integrated intensities, linewidths, decay constants, background contributions, and residual deviations from an assumed model \cite{Thompson1987,Ida2000,Olivero1977}. Peak positions may identify optical transitions, crystalline phases, vibrational modes, or chemical environments; linewidths may reflect instrumental resolution, disorder, homogeneous lifetime broadening, strain, finite-size effects, or microstructural heterogeneity; and integrated areas may be associated with relative populations, oscillator strengths, phase fractions, or signal contributions. In time-domain experiments, exponential or non-exponential decay parameters can similarly report recombination, relaxation, diffusion, or kinetic processes.

Despite its importance, nonlinear fitting remains a practical bottleneck in many laboratories. General-purpose numerical environments provide high flexibility but require users to construct model functions, manage initial guesses, impose bounds, choose weighting strategies, inspect residuals, assess convergence, examine parameter correlations, and manually export results. Conversely, many graphical tools are easy to use but restrict the user to a limited set of functions or hide numerical assumptions. This creates a gap between mathematically transparent, reproducible fitting and accessible daily analysis.

A further challenge is that a visually acceptable fit does not automatically imply that the inferred parameters are unique, stable, or scientifically interpretable. Overlapping peaks, highly correlated parameters, broad backgrounds, and flexible custom expressions can produce fit solutions in which several parameter combinations yield nearly indistinguishable model curves. Accordingly, a fitting environment intended for research use should not only return a best-fit curve, but also help the user inspect residuals, parameter uncertainties, parameter correlations, covariance structure, repeated-search stability, and the propagated uncertainty of quantities derived from fitted parameters.

\fited\ was developed to address these needs. It provides a graphical desktop workflow in which users can load two-column experimental data, choose the independent and dependent columns, crop a region of interest, select background and peak models, adjust parameter values and bounds, preview model components, execute weighted nonlinear least-squares fits, and export both fitted curves and parameter tables. The software also supports user-defined analytical profiles, enabling the same environment to be used for peak decomposition, decay analysis, and more specialized experiment-specific models. The current software release expands the original workflow with optimizer-comparison modes, Latin Hypercube Sampling for trial exploration, prominence-based peak detection, residual-guided missing-component suggestions, stability testing, batch fitting, session-level result histories, covariance-aware derived-quantity uncertainty propagation, and graphical uncertainty-contribution maps.

The application metadata in the manuscript uses the archived software DOI reported for the FitED software record \cite{AboulsaadFitEDZenodo}. The manuscript should be synchronized with the definitive DOI and version metadata of the distributed software package before final journal submission.

\section{Statement of need}

Several established tools support curve fitting or domain-specific peak analysis. Fityk is a widely used open-source program for nonlinear curve fitting and peak decomposition, especially for sums of bell-shaped functions in diffraction, chromatography, and spectroscopic data \cite{Wojdyr2010}. GSAS-II and FullProf provide sophisticated crystallographic workflows for diffraction data reduction, structure solution, and Rietveld refinement \cite{Toby2013,RodriguezCarvajal1993}. WinPLOTR is associated with powder diffraction pattern visualization and analysis within the FullProf ecosystem \cite{Roisnel2001}. Mantid provides a broad analysis and visualization framework for neutron scattering and muon spin experiments \cite{Arnold2014}. These packages demonstrate the importance of specialized scientific software for transforming raw measurements into interpretable parameters.

\fited\ is not intended to replace these mature tools. Its contribution is a lightweight and extensible fitting environment for one-dimensional, two-column experimental data when the researcher needs both routine peak-profile fitting and general analytical model fitting. In particular, \fited\ combines exact Voigt modeling, Pseudo-Voigt fitting, arbitrary custom analytical expressions, parameter-bound management, local and global-assisted optimizer options, automatic pre-fit search, residual-guided model refinement, fit-stability testing, covariance/correlation diagnostics, and reproducible session/export handling in one workflow. This combination is useful for researchers who move between different experimental modalities, for example fitting overlapping photoluminescence peaks in one dataset, comparing Raman or diffraction line-shape changes in another, and analyzing non-single-exponential decay traces in a third.

A specific design goal of \fited\ is to make statistical and numerical diagnostics visible without requiring the user to manually reconstruct them from code. The software therefore exposes parameter uncertainties, parameter correlations, optimizer selection summaries, repeated-search stability reports, and first-order covariance propagation for user-defined derived quantities. These features support a more reproducible practice of nonlinear fitting: the user can assess not only whether a curve passes through the data, but also whether the inferred quantities remain stable under alternative initializations and how uncertainty in fitted parameters propagates into scientifically reported observables.

\section{Software architecture and analysis workflow}

\fited\ is implemented as a layered Python application composed of a numerical backend and a desktop application layer. The backend contains reusable scientific-computing routines for data import, preprocessing, model construction, custom-expression validation, peak-profile definition, background modeling, optimization, automatic trial generation, peak detection, residual analysis, stability testing, uncertainty propagation, session normalization, and batch-fit output writing. The desktop layer provides file loading, parameter entry, visualization, dialogs, worker-thread orchestration, progress reporting, cancellation handling, result-history management, and ZIP-based export. This separation keeps mathematical fitting logic independent of interface-specific operations and makes the software easier to maintain, extend, and test.

\begin{figure}[t]
    \centering
    \includegraphics[width=0.98\linewidth]{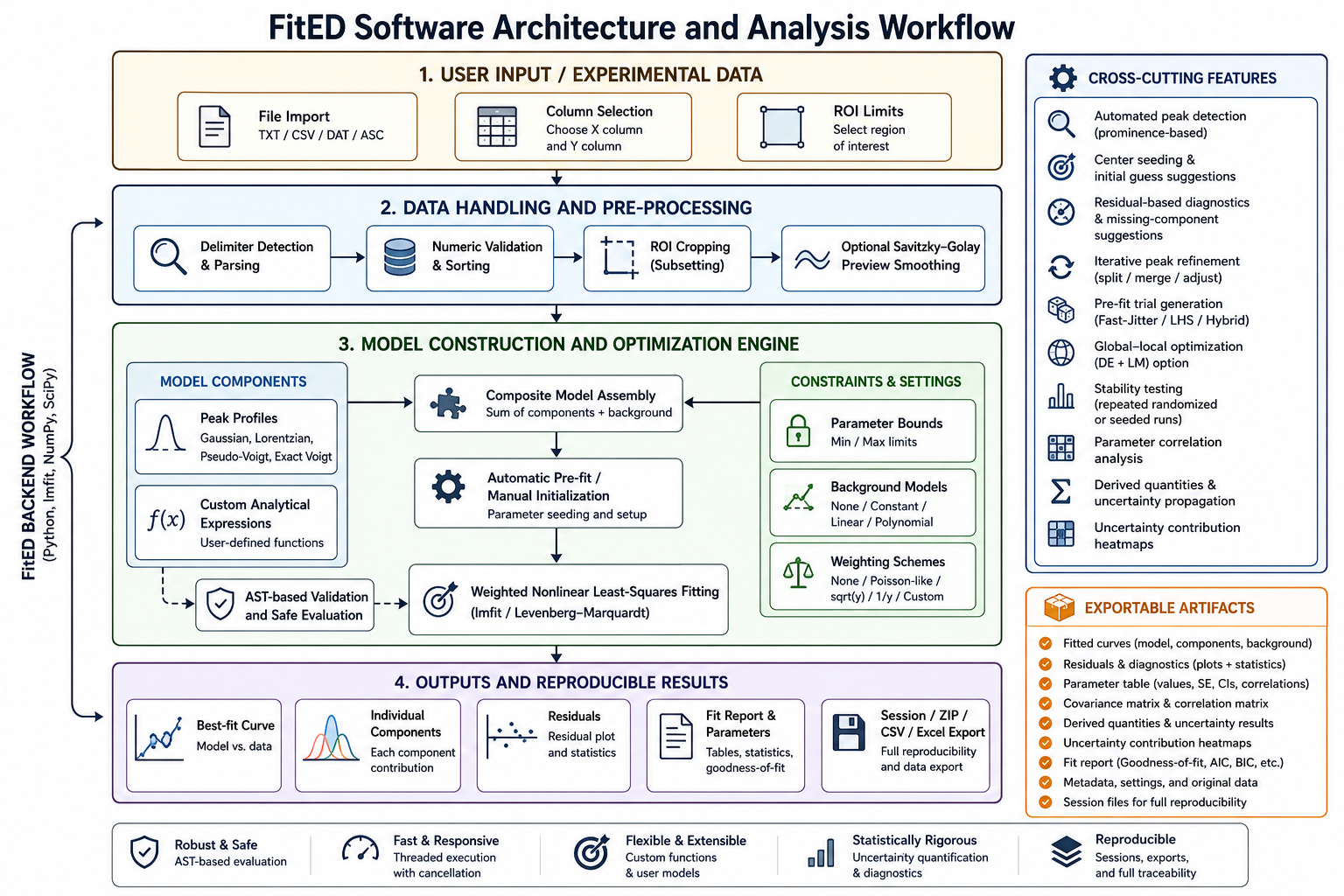}
    \caption{Software architecture and analysis workflow of \fited. Experimental data are imported, validated, optionally cropped and preview-smoothed, and passed to the model-construction and optimization engine. The fitting engine combines built-in peak profiles, custom analytical expressions, background models, parameter constraints, weighting strategies, optimizer comparison, automated trial sampling, residual-guided diagnostics, and uncertainty propagation to produce reproducible fitted curves, residuals, component contributions, fit reports, derived quantities, stability reports, and exportable session files.}
    \label{fig:fited_architecture}
\end{figure}

The architecture follows a data-analysis pipeline, illustrated in Fig.~\ref{fig:fited_architecture}. Experimental data are first imported from text-based files and converted into numerical arrays. The preprocessing layer validates numerical values, sorts data by the independent variable, applies the selected region of interest, and optionally generates a Savitzky--Golay smoothed trace for visual preview or selected detection routines only \cite{Savitzky1964}. The fitting layer constructs a composite model from selected peak profiles, custom analytical expressions, and background terms. Parameter bounds, weighting choices, initialization values, optimization mode, sampling mode, trial count, and optional random seed are then passed to the numerical fitting engine, which uses the \texttt{lmfit} modeling framework and SciPy-compatible minimization backends \cite{Newville2014,Virtanen2020}. The final layer exports fitted curves, individual components, residuals, fitted parameters, fit reports, optimizer summaries, metadata, session snapshots, derived quantities, propagated uncertainties, uncertainty-contribution heatmaps, and optional spreadsheet files.

At the implementation level, \fited\ uses the scientific Python ecosystem. NumPy provides numerical arrays and vectorized operations \cite{Harris2020}; pandas supports tabular input/output and result organization \cite{McKinney2010}; SciPy supplies numerical routines, including Savitzky--Golay filtering, exact Voigt evaluation, global-optimization support, and peak-finding operations \cite{Virtanen2020}; Matplotlib is embedded into the desktop application for visualization \cite{Hunter2007}; and \texttt{lmfit} supplies model objects, parameter containers, bounds, expression-constrained parameters, component evaluation, covariance estimates, and nonlinear least-squares fitting \cite{Newville2014}. The desktop application is implemented using Tkinter/ttk, and standalone executable generation is supported through PyInstaller-compatible packaging.

\section{Mathematical formulation}

\subsection{Composite model, weighted residual objective, and optimization strategy}

For a measured dataset $\{(x_i,y_i)\}_{i=1}^{N}$, \fited\ represents the fitted signal as an additive composite model,
\begin{equation}
    y_i = B(x_i;\vect{\beta}) + \sum_{j=1}^{N_p} P_j(x_i;\vect{\theta}_j) + \varepsilon_i,
    \label{eq:composite_data}
\end{equation}
where $B(x_i;\vect{\beta})$ is an optional background model, $P_j(x_i;\vect{\theta}_j)$ is the $j$th peak or custom component, $N_p$ is the number of active components, and $\varepsilon_i$ is the residual error. The full model prediction is
\begin{equation}
    f(x_i;\vect{\theta}) =
    B(x_i;\vect{\beta}) + \sum_{j=1}^{N_p} P_j(x_i;\vect{\theta}_j),
\end{equation}
where $\vect{\theta}$ collects all independently varied parameters and expression-defined dependent parameters stored in the model state.

The residual at data point $i$ is
\begin{equation}
    r_i(\vect{\theta}) = y_i - f(x_i;\vect{\theta}).
\end{equation}
When user-selected weights $w_i$ are supplied, \fited\ minimizes the weighted least-squares objective
\begin{equation}
    \chi^2(\vect{\theta})
    =
    \sum_{i=1}^{N}
    \left[w_i r_i(\vect{\theta})\right]^2
    =
    \left\|
    \mat{W}
    \left[
    \vect{y}-\vect{f}(\vect{\theta})
    \right]
    \right\|_2^2,
    \label{eq:chisq}
\end{equation}
where $\mat{W}=\mathrm{diag}(w_1,w_2,\ldots,w_N)$ is the diagonal weight matrix. The fitted parameter vector is therefore
\begin{equation}
    \widehat{\vect{\theta}}
    =
    \operatorname*{arg\,min}_{\vect{\theta}}
    \chi^2(\vect{\theta}),
    \label{eq:argmin}
\end{equation}
subject to lower and upper bounds set by the user, by parameter constraints, or by selected initialization strategies.

The reduced chi-square statistic reported by \fited\ is
\begin{equation}
    \chi^2_{\nu} = \frac{\chi^2}{\nu},
    \qquad
    \nu=N-k,
    \label{eq:redchi}
\end{equation}
where $k$ is the number of varied fitting parameters and $\nu$ is the number of residual degrees of freedom. In addition to $\chi^2$ and $\chi^2_\nu$, \fited\ can use the Akaike Information Criterion (AIC) and Bayesian Information Criterion (BIC) for trial or optimizer-candidate selection:
\begin{align}
    \mathrm{AIC} &= 2k - 2\ln\widehat{L},
    \label{eq:aic}\\
    \mathrm{BIC} &= k\ln N - 2\ln\widehat{L},
    \label{eq:bic}
\end{align}
where $\widehat{L}$ is the maximum likelihood. Lower values are favored when candidate solutions are compared under a common likelihood framework \cite{Akaike1974,Schwarz1978}.

In the default local optimizer mode, \fited\ uses the Levenberg--Marquardt (LM) route provided through \texttt{lmfit} \cite{Levenberg1944,Marquardt1963,More1978,Newville2014}. Let $\mat{J}$ be the residual Jacobian,
\begin{equation}
    J_{ij} =
    \frac{\partial r_i}{\partial \theta_j},
\end{equation}
where $J_{ij}$ measures how residual $r_i$ changes with parameter $\theta_j$. A Levenberg--Marquardt-type update solves
\begin{equation}
    \left(
    \mat{J}^{\mathrm{T}}\mat{J}
    +
    \lambda \mat{D}
    \right)
    \Delta\vect{\theta}
    =
    \mat{J}^{\mathrm{T}}\vect{r},
    \label{eq:lm_update}
\end{equation}
where $\lambda$ is a damping parameter, $\mat{D}$ is a diagonal scaling or damping matrix, and $\Delta\vect{\theta}$ is the proposed parameter update. Small $\lambda$ gives a Gauss--Newton-like step, whereas large $\lambda$ damps the update and improves stability for poorly conditioned local models.

For difficult spectra, poor initial guesses, or multi-parameter custom models, a local method may converge to a nearby local minimum rather than to the best available basin. \fited\ therefore provides a robust mode that compares a direct LM candidate with a Differential Evolution plus LM candidate. Differential Evolution (DE) is a population-based global optimizer for bounded continuous parameter spaces \cite{StornPrice1997}. For a population vector $\vect{x}_{i,g}$ at generation $g$, DE constructs a mutant vector
\begin{equation}
    \vect{v}_{i,g}
    =
    \vect{x}_{r_1,g}
    +
    F_{\mathrm{DE}}
    \left(
    \vect{x}_{r_2,g}
    -
    \vect{x}_{r_3,g}
    \right),
    \label{eq:de_mutation}
\end{equation}
where $r_1$, $r_2$, and $r_3$ are distinct population indices and $F_{\mathrm{DE}}$ is the mutation scale factor. Crossover then forms a trial vector $\vect{u}_{i,g}$ according to
\begin{equation}
    u_{i,g}^{(m)}
    =
    \begin{cases}
    v_{i,g}^{(m)}, & \text{if } q_m \leq C_R \text{ or } m=m_{\mathrm{rand}},\\
    x_{i,g}^{(m)}, & \text{otherwise},
    \end{cases}
    \label{eq:de_crossover}
\end{equation}
where $m$ indexes the parameter dimension, $q_m$ is a random number drawn from $[0,1]$, $C_R$ is the crossover probability, and $m_{\mathrm{rand}}$ ensures that at least one component is inherited from the mutant vector. Selection keeps the lower-objective candidate:
\begin{equation}
    \vect{x}_{i,g+1}
    =
    \begin{cases}
    \vect{u}_{i,g}, & \text{if } \chi^2(\vect{u}_{i,g}) \leq \chi^2(\vect{x}_{i,g}),\\
    \vect{x}_{i,g}, & \text{otherwise}.
    \end{cases}
    \label{eq:de_selection}
\end{equation}

In \fited, DE is not used as an unconditional replacement for LM. Instead, the robust optimizer mode evaluates two candidates:
\begin{enumerate}
    \item direct LM from the current user-defined or auto-seeded parameter state;
    \item DE global search followed by LM local polishing, denoted DE+LM.
\end{enumerate}
Each candidate is scored using the selected criterion, and the final robust-mode result is
\begin{equation}
    \widehat{\vect{\theta}}_{\mathrm{robust}}
    =
    \operatorname*{arg\,min}_{c\in\{\mathrm{LM},\,\mathrm{DE+LM}\}}
    S_c,
    \label{eq:robust_selection}
\end{equation}
where $S_c$ is the selected score for candidate $c$, such as $\chi^2$, $\chi^2_\nu$, AIC, or BIC. This comparison prevents a good local solution from being discarded simply because a stochastic global search did not find a better basin within the allocated search budget.

Because DE requires finite parameter bounds, \fited\ preserves user-defined finite bounds and generates conservative finite fallback bounds only for otherwise unbounded varied parameters during the global-search stage. The subsequent LM polish begins from the DE-derived parameter set and provides a locally refined solution. The selected optimizer candidate, candidate scores, criterion, and optional random seed are recorded in the fit report and export metadata.
\subsection{Built-in peak profiles}

\paragraph{Gaussian profile.}
For integrated area $A$, center $\mu$, and standard deviation $\sigma$, the area-normalized Gaussian component is
\begin{equation}
    G(x;A,\mu,\sigma) = \frac{A}{\sigma\sqrt{2\pi}}\exp\left[-\frac{(x-\mu)^2}{2\sigma^2}\right],
    \label{eq:gaussian}
\end{equation}
with
\begin{equation}
    \int_{-\infty}^{\infty}G(x;A,\mu,\sigma)\,dx = A.
\end{equation}
The Gaussian full width at half maximum is
\begin{equation}
    w_G = 2\sqrt{2\ln 2}\,\sigma \approx 2.354820045\,\sigma.
    \label{eq:gaussian_fwhm}
\end{equation}
Gaussian broadening is frequently associated with instrumental resolution, inhomogeneous disorder, or other approximately normally distributed perturbations \cite{Thompson1987,Ida2000}.

\paragraph{Lorentzian profile.}
For integrated area $A$, center $\mu$, and half width at half maximum $\gamma$, the normalized Lorentzian component is
\begin{equation}
    L(x;A,\mu,\gamma) = \frac{A}{\pi}\frac{\gamma}{(x-\mu)^2+\gamma^2}.
    \label{eq:lorentzian}
\end{equation}
It satisfies
\begin{equation}
    \int_{-\infty}^{\infty}L(x;A,\mu,\gamma)\,dx=A,
\end{equation}
with FWHM
\begin{equation}
    w_L = 2\gamma.
\end{equation}
Lorentzian line shapes are often used for homogeneous broadening, resonance profiles, and lifetime-related broadening mechanisms \cite{Olivero1977}.

\paragraph{Pseudo-Voigt profile.}
Many experimental features contain both Gaussian-like and Lorentzian-like broadening. The Pseudo-Voigt function approximates a Voigt-like line shape as a weighted combination of Gaussian and Lorentzian components \cite{Thompson1987,Ida2000}:
\begin{equation}
    P(x;A,\mu,w,\eta) = (1-\eta)G(x;A,\mu,w)+\eta L(x;A,\mu,w),
    \label{eq:pseudovoigt}
\end{equation}
where $0\leq \eta\leq 1$ is the Lorentzian mixing fraction in the conventional formulation. In \fited, the user-facing label explicitly clarifies this interpretation as \emph{L-fraction}, with $0$ indicating the Gaussian limit and $1$ indicating the Lorentzian limit.

\paragraph{Exact Voigt profile.}
The exact Voigt profile is the convolution of a Gaussian and a Lorentzian and is appropriate when both inhomogeneous and homogeneous broadening contribute to the observed line shape \cite{Olivero1977,Humlicek1982,Zaghloul2012}. Let
\begin{equation}
    G_0(x;\sigma)=\frac{1}{\sigma\sqrt{2\pi}}\exp\left(-\frac{x^2}{2\sigma^2}\right),
    \qquad
    L_0(x;\gamma)=\frac{1}{\pi}\frac{\gamma}{x^2+\gamma^2}.
\end{equation}
The unit-area Voigt profile centered at zero is
\begin{equation}
    V_0(x;\sigma,\gamma)=\int_{-\infty}^{\infty}G_0(x';\sigma)L_0(x-x';\gamma)\,dx'.
    \label{eq:voigt_convolution}
\end{equation}
Equivalently, it can be written using the real part of the Faddeeva function $w(z)$ \cite{Humlicek1982,Zaghloul2012}:
\begin{equation}
    V_0(x;\sigma,\gamma)=\frac{\operatorname{Re}[w(z)]}{\sigma\sqrt{2\pi}},
    \qquad
    z=\frac{x+i\gamma}{\sigma\sqrt{2}}.
\end{equation}
For integrated area $A$ and center $\mu$, \fited\ evaluates
\begin{equation}
    V(x;A,\mu,\sigma,\gamma)=A\,V_0(x-\mu;\sigma,\gamma),
    \label{eq:voigt_area}
\end{equation}
so that $\int_{-\infty}^{\infty}V(x;A,\mu,\sigma,\gamma)\,dx=A$. This area-normalized convention makes the fitted amplitude directly interpretable as integrated intensity. The Voigt FWHM has no simple closed-form expression, but the widely used Olivero--Longbothum approximation is \cite{Olivero1977}
\begin{equation}
    w_V \approx 0.5346\,w_L + \sqrt{0.2166\,w_L^2+w_G^2},
    \label{eq:olivero}
\end{equation}
where $w_G=2\sqrt{2\ln2}\,\sigma$ and $w_L=2\gamma$. The same relation is used by \fited\ when an approximate total width is needed for trial-generation logic or result interpretation.

\subsection{Backgrounds, weighting, and custom analytical profiles}

\fited\ supports no background, constant, linear, polynomial, and user-defined custom backgrounds,
\begin{align}
    B(x) &= 0, \\
    B(x) &= c, \\
    B(x) &= mx+b, \\
    B(x) &= \sum_{k=0}^{d}c_k x^k, \\
    B_{\mathrm{custom}}(x;\vect{\psi}) &= G(x,\psi_1,\psi_2,\ldots,\psi_n).
\end{align}
These additive backgrounds help account for detector offsets, sloping baselines, fluorescence continua, scattering backgrounds, drift, or slowly varying instrumental response. For custom backgrounds, the user supplies an analytical expression together with parameter names, default values, and bounds.

The implemented weighting options include unweighted fitting,
\begin{equation}
    w_i=1,
\end{equation}
Poisson-like weighting,
\begin{equation}
    w_i=\frac{1}{\sqrt{\max(|y_i|,1)}},
\end{equation}
square-root emphasis weighting,
\begin{equation}
    w_i=\sqrt{\max(|y_i|,\epsilon)},
\end{equation}
and inverse-intensity weighting,
\begin{equation}
    w_i=\frac{1}{\max(|y_i|,\epsilon)},
\end{equation}
where $\epsilon$ is a small positive value used to avoid division by zero. The \emph{sqrt($y$) emphasis} option intentionally gives larger objective-function weight to higher-intensity regions, whereas inverse-intensity weighting emphasizes weaker regions more strongly. Since weighting can change the fitted solution, \fited\ stores the selected weighting strategy in the session metadata.

Custom analytical profiles are represented as
\begin{equation}
    C(x;\vect{\phi}) = F(x,\phi_1,\phi_2,\ldots,\phi_m),
\end{equation}
where $F$ is supplied by the user. This allows models such as a monoexponential decay,
\begin{equation}
    I(t)=A\exp\left[-\frac{t-t_0}{\tau}\right]+c,
\end{equation}
a biexponential decay,
\begin{equation}
    I(t)=A_1\exp\left(-\frac{t}{\tau_1}\right)+A_2\exp\left(-\frac{t}{\tau_2}\right)+c,
\end{equation}
or a stretched exponential,
\begin{equation}
    I(t)=A\exp\left[-\left(\frac{t}{\tau}\right)^\beta\right]+c.
\end{equation}
The custom-expression system validates mathematical expressions before fitting by parsing the expression as an abstract syntax tree, restricting allowed syntax and functions, checking parameter names, and rejecting expressions that return arrays of the wrong shape or non-finite values. This mechanism permits high flexibility while reducing the risk of malformed expressions or unexpected execution behavior.

\section{Parameter covariance, correlations, and propagated uncertainty}

\subsection{Covariance matrix of fitted parameters}

After a successful local fit, the optimizer backend may estimate a covariance matrix for the independently varied parameters. Let
\begin{equation}
    \vect{\theta} = (\theta_1,\theta_2,\ldots,\theta_k)^{\mathrm{T}}
\end{equation}
be the vector of $k$ independently varied parameters. The associated covariance matrix is
\begin{equation}
    \mat{C} =
    \begin{bmatrix}
        \operatorname{Var}(\theta_1) & \operatorname{Cov}(\theta_1,\theta_2) & \cdots & \operatorname{Cov}(\theta_1,\theta_k) \\
        \operatorname{Cov}(\theta_2,\theta_1) & \operatorname{Var}(\theta_2) & \cdots & \operatorname{Cov}(\theta_2,\theta_k) \\
        \vdots & \vdots & \ddots & \vdots \\
        \operatorname{Cov}(\theta_k,\theta_1) & \operatorname{Cov}(\theta_k,\theta_2) & \cdots & \operatorname{Var}(\theta_k)
    \end{bmatrix}.
    \label{eq:covariance_matrix_shape}
\end{equation}
The diagonal terms define the parameter variances,
\begin{equation}
    C_{jj}=\operatorname{Var}(\theta_j),
\end{equation}
and therefore the reported parameter standard errors are
\begin{equation}
    s_{\theta_j}=\sqrt{C_{jj}}.
    \label{eq:param_stderr}
\end{equation}
The off-diagonal elements quantify pairwise co-variation of parameters in the local linearized approximation around the optimum. For standard least-squares theory, a common local approximation is
\begin{equation}
    \mat{C} \approx s^2 \left(\mat{J}^{\mathrm{T}}\mat{J}\right)^{-1},
    \label{eq:cov_approx}
\end{equation}
where $s^2$ is a residual-variance scale estimate and $\mat{J}$ is the residual Jacobian evaluated near the optimum. In practice, \fited\ uses the covariance estimate supplied by the \texttt{lmfit} fit result when it is available \cite{Newville2014}. Covariance-based diagnostics may be unavailable or unreliable if the Jacobian is ill-conditioned, if the covariance matrix cannot be estimated, or if the linear approximation is not informative for a strongly nonlinear solution.

\subsection{Correlation coefficients}

To make parameter coupling easier to interpret across parameters with different units and scales, \fited\ reports correlation coefficients derived from the covariance matrix. The correlation coefficient between parameters $\theta_i$ and $\theta_j$ is
\begin{equation}
    \rho_{ij} = \frac{C_{ij}}{\sqrt{C_{ii}C_{jj}}}.
    \label{eq:parameter_correlation}
\end{equation}
The corresponding correlation matrix is dimensionless and bounded elementwise by $-1\leq\rho_{ij}\leq 1$. Values near $+1$ or $-1$ indicate strong local linear coupling, meaning that alternative parameter combinations can vary together while producing similar local residual behavior. This is particularly important for overlapping peaks, broad backgrounds, area--width trade-offs, and center correlations in partially resolved components.

The exported fit report includes \texttt{lmfit}'s correlation reporting above a configurable threshold. In the current desktop workflow, correlations with magnitude at least $0.5$ are printed in the fit report to draw the user's attention to potentially important parameter dependencies.

\subsection{Derived quantities and first-order covariance propagation}

Researchers frequently report quantities that are not single fitted parameters but functions of several fitted parameters. Examples include
\begin{align}
    g_1 &= \frac{A_1}{A_2}, \\
    g_2 &= \frac{A_1}{A_1+A_2}, \\
    g_3 &= \mu_2-\mu_1, \\
    g_4 &= \frac{w_1}{w_2}.
\end{align}
\fited\ therefore allows the user to define a scalar derived quantity
\begin{equation}
    g = g(\vect{\theta})
\end{equation}
through a validated analytical expression that may reference fitted parameter names and approved mathematical functions. The scalar value is evaluated using the final fitted parameter set.

To propagate parameter uncertainty into $g$, \fited\ uses the first-order law of propagation of uncertainty \cite{JCGM100}. Let
\begin{equation}
    \nabla g =
    \begin{bmatrix}
        \partial g / \partial \theta_1 \\
        \partial g / \partial \theta_2 \\
        \vdots \\
        \partial g / \partial \theta_k
    \end{bmatrix}
    \label{eq:derived_gradient}
\end{equation}
be the gradient of the derived quantity with respect to the independently varied fitted parameters. The propagated variance is
\begin{equation}
    u_g^2 = \operatorname{Var}(g) \approx (\nabla g)^{\mathrm{T}}\mat{C}(\nabla g),
    \label{eq:propagation_matrix}
\end{equation}
and the propagated standard error reported by \fited\ is
\begin{equation}
    u_g = \sqrt{u_g^2}.
    \label{eq:absolute_uncertainty}
\end{equation}
Because $g$ may have units different from the raw fitted parameters, $u_g$ is an \emph{absolute uncertainty} expressed in the same units as the derived quantity itself. For interpretation, a relative uncertainty may be written as
\begin{equation}
    u_{g,\mathrm{rel}} = \frac{u_g}{|g|},
\end{equation}
when $g\neq 0$, although the present user-facing output emphasizes the absolute propagated standard error $u_g$.

For an explicit component-wise form, Eq.~\eqref{eq:propagation_matrix} becomes
\begin{equation}
    u_g^2 \approx
    \sum_{i=1}^{k}\left(\frac{\partial g}{\partial \theta_i}\right)^2 C_{ii}
    + 2\sum_{i=1}^{k}\sum_{j=i+1}^{k}
    \left(\frac{\partial g}{\partial \theta_i}\right)
    \left(\frac{\partial g}{\partial \theta_j}\right)
    C_{ij}.
    \label{eq:propagation_expanded}
\end{equation}
The first term contains contributions from individual parameter variances, whereas the second term contains covariance contributions. Neglecting off-diagonal covariance terms can substantially misrepresent the uncertainty of ratios, differences, linewidth combinations, or any derived observable involving correlated fit parameters.

\subsection{Numerical derivatives used in derived uncertainty propagation}

Rather than requiring a symbolic derivative for every possible user-defined expression, \fited\ evaluates the gradient numerically with finite differences. For parameter $\theta_j$ and step size $h_j$, a central-difference approximation is used whenever both perturbed points remain within bounds,
\begin{equation}
    \frac{\partial g}{\partial \theta_j}
    \approx
    \frac{g(\theta_j+h_j)-g(\theta_j-h_j)}{2h_j}.
    \label{eq:central_difference}
\end{equation}
If a bound prevents two-sided perturbation, the implementation falls back to an appropriate one-sided difference,
\begin{equation}
    \frac{\partial g}{\partial \theta_j}
    \approx
    \frac{g(\theta_j+h_j)-g(\theta_j)}{h_j}
\end{equation}
or
\begin{equation}
    \frac{\partial g}{\partial \theta_j}
    \approx
    \frac{g(\theta_j)-g(\theta_j-h_j)}{h_j}.
\end{equation}
The implementation updates expression-constrained dependent parameters before evaluating each perturbed derived expression. Consequently, derived quantities can reference both independently varied parameters and parameters defined in the fitting model through internal algebraic constraints. If derivatives cannot be evaluated, or if no covariance matrix is available, \fited\ preserves the derived value while issuing warnings that propagated uncertainty could not be fully determined.

\subsection{Uncertainty-contribution maps for derived quantities}

Beyond reporting a scalar propagated uncertainty $u_g$, \fited\ computes a parameter-resolved decomposition of the propagated variance. Define
\begin{equation}
    \vect{v} = \mat{C}\nabla g.
\end{equation}
The signed contribution associated with parameter $\theta_i$ is defined as
\begin{equation}
    q_i = \left(\frac{\partial g}{\partial \theta_i}\right) v_i
    = \left(\frac{\partial g}{\partial \theta_i}\right)
    \left[\mat{C}\nabla g\right]_i.
    \label{eq:uncertainty_contribution}
\end{equation}
Summing these contributions recovers the propagated variance,
\begin{equation}
    \sum_{i=1}^{k} q_i = (\nabla g)^{\mathrm{T}}\mat{C}(\nabla g) = u_g^2.
    \label{eq:contribution_sum}
\end{equation}
\fited\ visualizes the normalized signed percentage contribution
\begin{equation}
    Q_i = 100\times\frac{q_i}{u_g^2}.
    \label{eq:uncertainty_percent}
\end{equation}
For each derived quantity, the values $Q_i$ sum to approximately $100\%$ up to numerical precision. Positive values indicate variance-increasing contributions, whereas negative values indicate covariance-driven cancellation that reduces the final propagated variance. Contributions may exceed $100\%$ when compensated by negative contributions from correlated parameters. This map provides an interpretable bridge between a single propagated uncertainty and the underlying covariance structure, identifying which fitted parameters dominate the uncertainty budget of each derived observable.

\section{Automatic initialization, trial generation, and residual diagnostics}

\subsection{Center seeding from manual or detected positions}

A core practical challenge in nonlinear fitting is the sensitivity to initial guesses. \fited\ supports center seeding from manually chosen centers or automatically detected candidate peaks. Given proposed center positions $c_j$, the backend estimates local amplitudes, characteristic widths, and bounds by inspecting the selected fitting region and the spacing between nearby centers. For a given center $c_j$, the local signal value is sampled at the nearest data point. A characteristic spacing $s_j$ is estimated from the nearest neighboring center distance, or from a fraction of the total $x$ span when only one center is available. The initial full width at half maximum is then constructed approximately as
\begin{equation}
    w_j^{(0)} \sim \max\left(\frac{s_j}{2},\frac{x_{\max}-x_{\min}}{200}\right),
\end{equation}
with lower and upper width bounds generated from this initial scale. The initial area follows the approximate scaling
\begin{equation}
    A_j^{(0)} \sim h_j^{(0)}w_j^{(0)},
    \label{eq:area_seed}
\end{equation}
where $h_j^{(0)}$ is a local height estimate relative to the minimum signal level in the selected region. This strategy supplies physically plausible starting points while retaining user control through editable bounds.

\subsection{Fast-Jitter trial generation}

The original Auto pre-fit strategy in \fited\ is preserved as the default \emph{Fast Jitter} method. Starting from seeded peak definitions, trial $t$ perturbs centers, widths, and amplitudes within controlled windows. In schematic form,
\begin{align}
    c_j^{(t)} &\leftarrow c_j + \delta c_j^{(t)}, \\
    w_j^{(t)} &\leftarrow w_j^{(0)}\,\alpha_w^{(t)}, \\
    A_j^{(t)} &\leftarrow A_j^{(0)}\,\alpha_A^{(t)},
\end{align}
where $\delta c_j^{(t)}$ is a bounded random center displacement and $\alpha_w^{(t)}$, $\alpha_A^{(t)}$ are randomized multiplicative factors drawn around discrete trial-scale choices. The actual implementation adaptively updates parameter bounds together with trial values, retaining finite widths and positive area bounds for standard positive-peak initialization. Fast Jitter is computationally inexpensive and remains useful when the current initial model is already close to a reasonable solution.

\subsection{Latin Hypercube Sampling and hybrid sampling}

For broader exploration of initialization space, \fited\ provides Latin Hypercube Sampling (LHS) \cite{McKay1979}. LHS stratifies each sampled dimension into $n$ intervals for $n$ trials, then combines independently permuted interval selections across dimensions. In one dimension, trial $t$ samples
\begin{equation}
    u_t = \frac{t+\xi_t}{n},
    \qquad \xi_t\sim U(0,1),
\end{equation}
followed by a random permutation of the $u_t$ values for each dimension. This construction improves marginal coverage of each parameter direction compared with purely independent random draws.

For standard peak profiles, \fited\ samples center, FWHM, and amplitude dimensions. For exact Voigt components, center, $\sigma$, $\gamma$, and amplitude are sampled separately. Trial values are transformed from $u\in[0,1]$ to finite parameter bounds by linear maps,
\begin{equation}
    z = z_{\min}+u\left(z_{\max}-z_{\min}\right),
    \label{eq:lhs_linear}
\end{equation}
or by logarithmic maps for strictly positive parameters spanning multiple orders of magnitude,
\begin{equation}
    z = \exp\left[\ln(z_{\min}) + u\left(\ln(z_{\max})-\ln(z_{\min})\right)\right].
    \label{eq:lhs_log}
\end{equation}
Custom parameters use the same finite-bound logic and switch to logarithmic sampling when both bounds are non-negative and a positive scale is appropriate.

A third mode, \emph{Hybrid: Fast Jitter + Latin Hypercube}, uses Fast-Jitter trials for the first portion of the trial budget and LHS trials for the remainder. This balances rapid local exploration around existing guesses with more systematic coverage of the wider admissible initialization space.

\subsection{Auto pre-fit search and trial selection}

For each generated trial, \fited\ constructs the full model, executes the selected optimizer mode, evaluates the chosen selection criterion, and stores the best finite result. If $S_t$ is the score of trial $t$, the retained candidate satisfies
\begin{equation}
    t^{\star} = \operatorname*{arg\,min}_{t}\; S_t,
\end{equation}
where $S_t$ may be AIC, BIC, chi-square, or reduced chi-square. The trial-generation method, criterion, optimizer mode, and optional seed are recorded for reproducibility. For custom-profile auto pre-fit, custom parameter values are randomized within user-defined bounds using the selected sampling approach.

\subsection{Find-peaks proposal workflow}

\fited\ includes an optional \emph{Find peaks} workflow intended to propose candidate peak centers before fitting. It applies one-dimensional peak detection to either the raw fitting signal or an optional Savitzky--Golay-smoothed detection trace. For positive peaks, candidate maxima are detected directly; for negative dips, the detection trace is sign-inverted before searching. The backend uses a prominence criterion expressed as a user-defined percentage of the signal range,
\begin{equation}
    p_{\min} = \frac{P_{\%}}{100}\left(y_{\max}^{\mathrm{det}}-y_{\min}^{\mathrm{det}}\right),
\end{equation}
where $P_{\%}$ is the selected prominence percentage and $y^{\mathrm{det}}$ is the signal used for detection. Minimum peak separation and minimum peak width are entered by the user in $x$ units and internally converted into sample counts using the median sampling step. The detected candidates are ranked by prominence, truncated to the user-specified maximum number of proposals, sorted by center position, and then converted into default Pseudo-Voigt peak definitions. \fited\ subsequently reuses the center-seeding routine to estimate areas, widths, and bounds before showing the candidate model table.

For negative-peak detection, the proposal logic converts amplitude guesses into negative values and sets compatible negative amplitude bounds. The Find-peaks feature is therefore a seeding helper rather than an autonomous physical assignment of peaks; the user remains responsible for deciding whether proposed components are scientifically meaningful.

\subsection{Residual-based missing-component suggestions}

After a fit, \fited\ can search the residual signal
\begin{equation}
    e_i = y_i - \widehat{f}(x_i)
\end{equation}
for structured positive or negative features that may indicate underfitting. The residual may optionally be smoothed for detection only. To avoid suggesting tiny noise fluctuations, the prominence threshold combines an absolute fraction of the original signal span with a robust residual-noise scale. The robust scale estimate is computed from the median absolute deviation (MAD),
\begin{equation}
    \widehat{\sigma}_{\mathrm{MAD}} = 1.4826\,\mathrm{median}\left(|e_i-\mathrm{median}(e)|\right),
    \label{eq:mad_sigma}
\end{equation}
which estimates the standard deviation under approximately normal residual noise and remains less sensitive to occasional large deviations than the ordinary standard deviation \cite{Hampel1974}. For the selected sensitivity level, \fited\ uses a threshold of the form
\begin{equation}
    T = \max\left(\alpha\,\Delta y,\,\beta\,\widehat{\sigma}_{\mathrm{MAD}}\right),
    \label{eq:residual_threshold}
\end{equation}
where $\Delta y$ is the raw data range and $(\alpha,\beta)$ are sensitivity-dependent constants. Conservative, Normal, and Aggressive settings correspond to progressively lower thresholds.

Residual candidates are filtered by the same distance and optional width controls used in peak detection, checked against currently active peak centers to avoid duplicate suggestions, ranked by prominence, and presented to the user for acceptance or rejection. Accepted suggestions are appended to the peak table as seeded Pseudo-Voigt components and may then be refined through the dedicated \emph{Refine with added peaks} workflow.

\subsection{Refine with added peaks}

The refinement workflow addresses the common situation in which an initial fit captures major components but leaves shoulders or weak residual structure. Let the first-stage model contain $N_{\mathrm{old}}$ active components and let the user append $N_{\mathrm{new}}$ additional candidate components. \fited\ copies fitted parameter values from the previous result for the old components, uses newly seeded values for the additional components, executes repeated refinement trials, and finally performs a full fit from the best intermediate candidate. In practice, this workflow stabilizes incremental model building: major components are not discarded when new residual-informed components are explored, while the final full fit still permits all active parameters to be jointly optimized within their bounds.

\section{Repeated stability testing}

A low residual or attractive fit statistic does not prove that the decomposition is numerically stable. To address this, \fited\ includes a repeated-search stability test. The user chooses one of two protocols:
\begin{enumerate}
    \item \emph{Repeat Run fit}: repeat the current fit protocol from the current model state;
    \item \emph{Repeat Auto pre-fit}: repeat the complete Auto pre-fit search protocol, including trial sampling, before selecting the best result for each repeat.
\end{enumerate}
The user specifies the number of repeats and a near-best tolerance $\Delta S$. If a base random seed $S_0$ is supplied, repeat $r$ uses
\begin{equation}
    S_r = S_0 + r
\end{equation}
modulo the admissible 32-bit seed range. If no seed is supplied, repeats remain stochastic.

For repeat $r$, the selected fit criterion produces a score $S_r^{\mathrm{fit}}$. Among successful repeats, \fited\ reports the best, mean, median, standard deviation, and worst score. The best score is
\begin{equation}
    S_{\min} = \min_r S_r^{\mathrm{fit}}.
\end{equation}
Near-best records are defined by
\begin{equation}
    S_r^{\mathrm{fit}} \leq S_{\min}+\Delta S.
    \label{eq:near_best}
\end{equation}
The stability report also records the frequency with which each optimizer candidate is selected in robust comparison mode and summarizes the distribution of fitted parameter values among near-best solutions. For parameter $\theta_j$, this includes observed minimum, maximum, mean, standard deviation, and spread across near-best solutions. A narrow spread indicates that the parameter is reproducibly recovered under the explored perturbations, whereas a broad spread warns that multiple parameter combinations are nearly competitive according to the selected score.

The stability test is intentionally presented as a diagnostic rather than proof of physical uniqueness. A stable repeated-search score does not guarantee that the model is scientifically correct, just as a wide spread does not automatically invalidate a physically motivated model. \fited\ therefore encourages joint interpretation of stability reports, residuals, correlations, propagated uncertainties, and domain-specific constraints.

\section{Graphical User Interface and implemented workflow}

The \fited\ graphical interface is designed to follow the natural sequence of a fitting workflow: data import, fit settings, component definition, fitting actions, reports, and visualization. The interface is divided into a tabbed control panel and an embedded plotting panel, allowing users to move from file loading to parameter adjustment, fitting, residual inspection, uncertainty analysis, and export within a single environment.

\subsection{Load data tab}

The data-loading controls allow users to open text-based files, select the delimiter, skip metadata rows, and choose independent and dependent variable columns. The current file can be reloaded with updated parsing settings. A reset action clears peak and fit state while preserving the current manually selected region of interest when desired. Imported data are plotted immediately in the embedded Matplotlib pane.

\subsection{Fit settings tab}

The fit-setting controls include the region of interest, optional Savitzky--Golay preview smoothing, background selection, polynomial order, weighting strategy, auto-fit trial count, optimizer mode, maximum number of fit evaluations, model-selection criterion, auto-fit sampling method, and optional random seed. Background choices include none, constant, linear, polynomial, and custom analytical backgrounds. When a custom background is selected, the interface exposes a profile selector and dynamically generated parameter fields.

Additional diagnostic controls include:
\begin{itemize}
    \item \emph{Find peaks}: maximum number of proposed peaks, prominence percentage, minimum distance, minimum width, direction, and optional smoothed detection;
    \item \emph{Residual missing peaks}: maximum number of residual suggestions, sensitivity level, direction, and optional smoothed residual detection.
\end{itemize}
These settings make automated proposal routines configurable rather than opaque.

\subsection{Peaks tab}

Users define one or more active components and assign each component a built-in or custom model. Built-in models include Gaussian, Lorentzian, Pseudo-Voigt, and exact Voigt profiles. For each component, the interface exposes parameter values and bounds for center, area, width, and model-specific quantities such as the Pseudo-Voigt Lorentzian fraction or exact Voigt $\sigma$ and $\gamma$. The custom-profile manager allows the user to define named analytical profiles through parameter tables and validated expressions; parameter entry fields are then generated dynamically in the active peak rows.

\subsection{Actions tab}

The Actions tab exposes the main fitting workflow:
\begin{itemize}
    \item Pick centers from plot;
    \item Find peaks;
    \item Auto pre-fit;
    \item Suggest missing peaks from residual;
    \item Refine with added peaks;
    \item Run fit;
    \item Fit stability test;
    \item Derived quantities;
    \item Cancel running fit;
    \item Save/load session files;
    \item Batch fit folder;
    \item Save ZIP results.
\end{itemize}
Long-running fit operations execute in worker threads so that the desktop interface remains responsive. A progress bar and status line summarize the current worker state, and the cancel control requests termination at safe checkpoints between fit calls or repeated trials.

The \emph{Derived quantities} action opens a post-fit dialog in which the user writes expressions of the form
\begin{equation*}
    \texttt{Name = expression}.
\end{equation*}
The dialog lists available fitted parameters, computes each derived value, reports propagated absolute standard uncertainty when covariance information is available, stores a textual report in the Reports tab, and automatically opens the derived uncertainty-contribution map described in Eqs.~\eqref{eq:uncertainty_contribution}--\eqref{eq:uncertainty_percent}.

\subsection{Reports tab and temporary result packages}

\fited\ preserves session-only histories for generated fit reports, stability-test reports, selected best-fit stability reports, derived-quantity reports, and temporary exportable result packages. Closing a popup therefore does not remove its text from the current application session. Each fitted result package can later be selected and saved as an explicit ZIP archive.

A result package can include fitted curves, residuals, component contributions, parameter tables, fit summary metadata, fit report text, session JSON snapshot, original source-file copy when available, derived quantities, uncertainty-contribution matrix data, uncertainty heatmap images, and spreadsheet output when \texttt{openpyxl} is available. Derived quantities and their uncertainty map are attached to the active result package only after they are computed for that specific fit result, preventing stale derived outputs from being silently carried into later fits.

\subsection{Batch fitting}

The current release supports folder-level batch fitting from the Actions tab. The user chooses an input folder, an output folder, file patterns, optional recursive scanning, and one of two modes:
\begin{enumerate}
    \item Run fit using current parameters as a template;
    \item Auto pre-fit then final fit for each file.
\end{enumerate}
The batch worker applies the selected template settings, records successful and failed files, writes CSV and optional spreadsheet summaries, and stores the batch template session JSON. This workflow supports repeated analysis of related spectra while preserving a clear statement that all spectra should reasonably share the chosen ROI, model structure, bounds, and fitting assumptions.

\section{Advanced statistical diagnostics and uncertainty analysis}

The current release of \fited\ extends the original fitting framework with a substantially expanded set of statistical diagnostics and uncertainty-analysis tools. In nonlinear fitting problems involving overlapping peaks, correlated parameters, broad backgrounds, constrained expressions, or custom analytical functions, a visually acceptable fit is not sufficient to guarantee that the extracted parameters are physically meaningful, numerically stable, or statistically identifiable.

The covariance matrix, parameter correlations, derived quantities, covariance-aware uncertainty propagation, numerical derivative evaluation, and uncertainty-contribution heatmaps are formulated mathematically in Section~5. The present section focuses on the additional diagnostic tools implemented in the current release and on the scientific interpretation of the exported diagnostics. Residual analysis, residual autocorrelation, and Q--Q plots are standard regression diagnostics for assessing systematic model error, residual independence, and approximate normality assumptions \cite{Montgomery2012}.

These tools are intended to answer complementary scientific questions:

\begin{itemize}
    \item Is the fit numerically stable?
    \item Are the parameters uniquely determined?
    \item Are some parameters strongly degenerate?
    \item Does the model systematically fail in some spectral region?
    \item Are the residuals statistically consistent with the assumptions underlying least-squares fitting?
    \item Which fitted parameters dominate the uncertainty of a reported observable?
    \item Are propagated uncertainties physically trustworthy?
\end{itemize}

Unlike traditional fitting workflows that expose only fitted parameter values and a scalar goodness-of-fit metric, \fited\ attempts to expose the local geometry of the nonlinear optimization landscape itself.

\subsection{Covariance ellipses and local parameter-space geometry}

The covariance ellipses implemented in \fited\ are derived from the local covariance approximation obtained from the fitted parameter covariance matrix. These ellipses therefore visualize the local uncertainty geometry and parameter-correlation structure around the optimum solution.

For two fitted parameters,
\begin{equation}
\vect{p} =
\begin{bmatrix}
\theta_1 \\
\theta_2
\end{bmatrix},
\end{equation}
their two-dimensional covariance submatrix is
\begin{equation}
\mat{C}_{2\mathrm{D}} =
\begin{bmatrix}
\sigma_1^2 & C_{12} \\
C_{21} & \sigma_2^2
\end{bmatrix},
\end{equation}
where $\sigma_1$ and $\sigma_2$ are the standard uncertainties of $\theta_1$ and $\theta_2$, respectively, and $C_{12}$ and $C_{21}$ are covariance terms between the two parameters.

The covariance ellipse is defined by
\begin{equation}
(\vect{p}-\widehat{\vect{p}})^{\mathrm T}
\mat{C}_{2\mathrm{D}}^{-1}
(\vect{p}-\widehat{\vect{p}})
=
k^2,
\end{equation}
where $\widehat{\vect{p}}$ is the optimal fitted parameter pair and $k$ controls the ellipse scale. The ellipse orientation is determined by the eigenvectors of $\mat{C}_{2\mathrm{D}}$, while the semi-axis lengths are
\begin{equation}
a_i = k \sqrt{\lambda_i},
\end{equation}
where $\lambda_i$ are the eigenvalues of the covariance submatrix.

The covariance ellipse answers the scientific question:
\begin{equation*}
\text{``How can two parameters vary together while preserving acceptable fit quality?''}
\end{equation*}

This visualization is particularly useful for overlapping peaks, center degeneracy, amplitude--width trade-offs, and broad background correlations. Strongly elongated ellipses indicate severe local parameter degeneracy and weak identifiability.

\subsection{Residual diagnostics}

Residual diagnostics analyze the fitted residual vector
\begin{equation}
r_i = y_i - f(x_i),
\end{equation}
where $y_i$ is the measured signal and $f(x_i)$ is the fitted model evaluated at the same coordinate.

The residual should ideally behave as approximately structureless random noise. Residual diagnostics therefore answer the scientific question:
\begin{equation*}
\text{``Is the model systematically wrong?''}
\end{equation*}

Good residuals should fluctuate randomly around zero, exhibit approximately constant variance, and avoid long-range oscillatory structure. Structured residuals may indicate missing peaks, incorrect background assumptions, unsuitable line-shape models, detector artifacts, or physically incomplete models.

\subsection{Residual autocorrelation}

The residual autocorrelation at lag $k$ is
\begin{equation}
\rho_k =
\frac{
\sum_{i=1}^{N-k}(r_i-\bar{r})(r_{i+k}-\bar{r})
}{
\sum_{i=1}^{N}(r_i-\bar{r})^2
},
\end{equation}
where $r_i$ are residuals, $\bar{r}$ is the residual mean, and $k$ is the lag index.

Residual autocorrelation answers the question:
\begin{equation*}
\text{``Do neighboring residuals remain statistically correlated?''}
\end{equation*}

For approximately random residuals, $\rho_k$ should fluctuate near zero. Strong positive or negative autocorrelation indicates systematic model mismatch rather than purely stochastic noise.

\subsection{Quantile--quantile residual analysis}

The quantile--quantile (Q--Q) residual plot compares empirical residual quantiles against theoretical Gaussian quantiles. If the residual distribution approximately follows
\begin{equation}
r_i \sim \mathcal{N}(0,\sigma_r^2),
\end{equation}
where $\sigma_r$ is the residual standard deviation, the ordered residual quantiles should approximately align along a straight line.

The Q--Q analysis answers the question:
\begin{equation*}
\text{``Are the residuals approximately Gaussian?''}
\end{equation*}

This is important because covariance-based uncertainty estimates rely on approximately Gaussian local residual behavior. Significant curvature or heavy-tail deviations in the Q--Q plot may indicate unresolved hidden peaks, incorrect weighting, detector saturation, or non-Gaussian experimental noise.

\subsection{Relationship between the diagnostics}

The diagnostics implemented in \fited\ are mathematically connected but answer different scientific questions:

\begin{table}[h]
\centering
\begin{tabular}{ll}
\toprule
Diagnostic & Primary scientific question \\
\midrule
Residual diagnostics & Is the model systematically wrong? \\
Residual autocorrelation & Are residuals sequentially correlated? \\
Q--Q plot & Are residuals approximately Gaussian? \\
Covariance matrix & What is the local uncertainty geometry? \\
Correlation matrix & Which parameters are strongly coupled? \\
Covariance ellipse & How do two parameters jointly vary? \\
Parameter uncertainty & How precisely is each parameter determined? \\
Derived quantities & What physical observable is computed? \\
Error propagation & How uncertain is the derived observable? \\
Heatmaps & Which parameters dominate uncertainty or correlation? \\
\bottomrule
\end{tabular}
\caption{Relationship between the advanced statistical diagnostics implemented in \fited.}
\end{table}

These diagnostics become particularly important for overlapping photoluminescence peaks, Raman multiplets, diffraction peak overlap, broad defect emission, constrained models, custom analytical functions, strongly correlated backgrounds, and Voigt/Pseudo-Voigt decomposition problems. In such systems, a visually good fit may still contain non-unique parameter sets, hidden covariance degeneracy, underestimated uncertainties, or systematic model mismatch.

The extended statistical framework implemented in \fited\ therefore transforms the fitting workflow from simple curve matching into a more rigorous uncertainty-aware parameter inference environment.

\section{Exported diagnostics and reproducibility}

\fited\ exports more than a fitted curve. The ZIP-based result archive can contain:
\begin{itemize}
    \item data, best-fit curve, residuals, and all evaluated component curves;
    \item fitted parameter values, standard errors where available, bounds, variation flags, and expression constraints;
    \item summary metadata including source file, ROI, weighting mode, fit criterion, optimizer mode, selected optimizer candidate, random seed, fit statistics, AIC, BIC, and function-evaluation count;
    \item a full fit report including \texttt{lmfit} statistics and parameter correlations;
    \item the session JSON payload needed to reconstruct fitting settings and custom profile definitions;
    \item derived quantity tables including value, propagated absolute standard uncertainty, variance, used parameters, and warnings;
    \item uncertainty-contribution matrix CSV output and the associated heatmap PNG;
\item correlation-matrix reports and correlation heatmap diagnostics;
\item residual-diagnostics summaries including residual autocorrelation analysis and Q--Q residual analysis;
\item covariance-ellipse reports and exported covariance-ellipse visualizations;
    \item an Excel workbook collecting the same tabular outputs when supported;
    \item a copy of the original spectrum file when the saved source path remains available.
\end{itemize}
This export strategy supports reproducibility because the numerical result, optimization choices, uncertainty calculations, and model state can be archived together rather than being reconstructed manually from screenshots or copied parameter values.

\section{Impact and research applications}

The primary impact of \fited\ is methodological. It provides a reproducible and user-accessible route for extracting quantitative parameters from experimental data containing overlapping features, broad backgrounds, or experiment-specific analytical responses. Many scientific questions in spectroscopy, photoluminescence, diffraction, and time-resolved measurements depend on reliable estimates of peak positions, linewidths, integrated intensities, background contributions, and decay parameters. These quantities are commonly obtained through nonlinear model fitting rather than direct inspection of raw data \cite{Thompson1987,Ida2000,Olivero1977}.

A first class of research questions enabled by \fited\ concerns systematic comparison of peak-shape parameters across related samples or experimental conditions. Researchers can compare fitted centers, areas, linewidths, and Gaussian/Lorentzian contributions as a function of composition, processing temperature, excitation condition, irradiation dose, aging time, or measurement environment. Since \fited\ supports Gaussian, Lorentzian, Pseudo-Voigt, and exact Voigt models, it can be used to examine whether a spectral or diffraction feature is better described by predominantly Gaussian broadening, Lorentzian broadening, mixed broadening, or a full convolutional Voigt form \cite{Olivero1977,Humlicek1982,Zaghloul2012}.

A second class of research applications involves decomposition of complex signals into physically interpretable components. In photoluminescence and optical spectroscopy, overlapping emission or absorption bands may originate from excitonic transitions, defect states, trap-assisted recombination pathways, surface states, or phase contributions. In diffraction and scattering data, overlapping peaks may reflect coexisting phases, structural distortions, strain, or broad size distributions. \fited\ improves the pursuit of these questions by allowing users to build multi-component models, constrain parameters, inspect fitted components, evaluate residuals, and use stability/uncertainty diagnostics to identify when decompositions are numerically fragile.

A third impact arises from custom analytical functions. Time-resolved photoluminescence, pump--probe spectroscopy, relaxation measurements, kinetic traces, and empirical response curves may require monoexponential, multiexponential, stretched-exponential, threshold, saturation, or other custom models. \fited\ allows such expressions to be defined directly and fitted using the same workflow. This enables researchers to address questions involving carrier recombination, relaxation dynamics, kinetic heterogeneity, or non-single-exponential behavior without modifying the source code.

A fourth impact is the treatment of \emph{reported observables} rather than only fitted parameters. Researchers often publish ratios of areas, differences between centers, linewidth ratios, normalized fractions, or more complicated scalar quantities derived from several fit parameters. \fited\ computes such values directly from validated expressions and propagates covariance-aware uncertainty into them. The uncertainty-contribution map then reveals whether the propagated uncertainty is dominated by one amplitude, one center, one linewidth, or by covariance cancellation between correlated parameters. This helps the user distinguish between a scientifically useful derived observable and a numerically unstable quantity that depends on poorly resolved parameter combinations.

Finally, the software improves daily fitting practice by reducing ad hoc scripting and by preserving model assumptions. Researchers often fit multiple spectra or traces while manually recording parameters and exporting figures. Such workflows are difficult to reproduce if initial guesses, bounds, weighting choices, optimizer choices, seeds, and background models are not preserved. \fited\ addresses this by storing the fitting state, custom profiles, fitted outputs, reports, stability diagnostics, derived-quantity outputs, and metadata. At the present stage, \fited\ is newly released; therefore, its impact is best described as methodological and near-term rather than based on established external citation counts. Future impact can be assessed through repository activity, archived release downloads, and publications that cite the software release \cite{AboulsaadFitEDZenodo}.

\section{Conclusion}

\fited\ provides an integrated environment for nonlinear fitting of one-dimensional scientific data. Its main contribution is the combination of physically meaningful peak profiles, exact Voigt evaluation, arbitrary user-defined analytical functions, weighted least-squares fitting, robust optimizer comparison, automated initialization search, residual-guided refinement, repeated stability testing, covariance/correlation diagnostics, derived-quantity uncertainty propagation, and reproducible export in a graphical desktop workflow. The software is designed for researchers who require ease of use while retaining detailed control over model assumptions, parameter bounds, and uncertainty interpretation.

The built-in Gaussian, Lorentzian, Pseudo-Voigt, and exact Voigt functions cover common spectroscopy and diffraction use cases, while the custom-profile architecture extends the same workflow to decay curves, empirical response functions, and specialized analytical models. By validating custom expressions, comparing optimizer candidates, stratifying trial exploration through Latin Hypercube Sampling, preserving sessions, diagnosing residual structures, computing propagated uncertainties for derived observables, and exporting complete results, \fited\ promotes transparent and reproducible fitting practice. Future development may include global fitting across multiple datasets, bootstrap and profile-likelihood uncertainty analysis, Bayesian posterior sampling, model-comparison visualizations, instrument-response convolution, and a plugin system for community-contributed models.

\section*{Acknowledgements}
The author gratefully acknowledges the scientific Python community for developing and maintaining the open-source libraries on which this software is built. The author also thanks Ahmed Abdelmagid, Zhe Zhang, and Jiaxin Guo for providing data used in testing. In addition, the author is grateful to Ahmed Abdelmagid, Enas Abd Abdelghafar, and Donia Ibrahim for their valuable discussions and feedback during the development of the software prior to its release.

\end{document}